\documentclass{article}
\usepackage{a4wide}
\usepackage{amsmath,amsthm,amssymb,amsfonts}
\usepackage[T1]{fontenc}
\usepackage{mathptmx}
\usepackage[pdftex,pdftitle={The role of the input scale in parton distribution analyses}, pdfauthor={P. Jimenez--Delgado}, colorlinks=true, linkcolor=black, citecolor=black, urlcolor=black]{hyperref}
\usepackage[all]{hypcap}
\usepackage{graphics,graphicx}
\usepackage{rotating,subfigure}
\usepackage{float}
\usepackage[numbers,sort&compress]{natbib}
\usepackage[small]{caption}
\newcommand{\as}{$\alpha_s(M_Z^2)$~}
\newcommand{\qs}{$Q_o^2$~}
\newcommand{\chisq}{$\chi^2$}

\begin{document}
\thispagestyle{empty}
\begin{center}
\begin{minipage}{\textwidth}
\begin{flushright}
ZU-TH 07/12\\
JLAB-THY-12-1578\\
Phys.~Lett.~{\bf B714} (2012) 301\\
\end{flushright}
\vspace{6mm}
\begin{center}
{\Huge The role of the input scale in\\ parton distribution analyses}\\[10mm]
{\Large P. Jimenez-Delgado}\\[5mm]
\textit{Institut f\"ur Theoretische Physik, Universit\"at Z\"urich, CH-8057 Z\"urich, Switzerland}\\
\textit{Thomas Jefferson National Accelerator Facility, VA 23606 Newport News, USA}\\[0.5em]
\end{center}
\vspace{3mm}\hspace{13.4mm}
\parbox{0.8\textwidth}{\small
A first systematic study of the effects of the choice of the input scale in global determinations of parton distributions and QCD parameters is presented. It is shown that, although in principle the results should not depend on these choices, in practice a relevant dependence develops as a consequence of what is called \emph{procedural bias}. This uncertainty should be considered in addition to other theoretical and experimental errors, and a practical procedure for its estimation is proposed. Possible sources of mistakes in the determination of QCD parameter from parton distribution analysis are pointed out.}
\end{minipage}
\end{center}
\vspace{6mm}

Parton distribution functions (PDFs) describe the nucleon in high-energy collisions and, in addition to their intrinsic interest, play an essential role in the analysis and understanding of scattering experiments, for example the ones currently performed at the LHC. For this reason a considerable effort is being continuously made in order to determine the PDFs as accurately as possible. During the last few years a number of groups have produced publicly available PDFs using different data sets and analysis frameworks, and some joint initiatives aiming to improve the precision and understanding of PDF determinations in various aspects have also been established \cite{Alekhin:2011sk,Alekhin:2010dd}.

As is well known, the $Q^2$ scale\footnote{We set equal renormalization and factorization scales, as has been done in all PDF analyses so far.} evolution of the PDFs $f(x,Q^2)$ is calculable within perturbative QCD and given by the renormalization group equations (RGE). In parton distribution analyses the $x$-dependence of the PDFs is thus extracted at a particular scale $Q_o^2$, usually referred to as the \emph{input} scale. One aspect of PDF determinations which has not been \emph{systematically}\footnote{Less systematic variations have been considered previously, e.g. in \cite{Gluck:2007ck,JimenezDelgado:2008hf} or by the HERAPDF collaboration (cf. \cite{Aad:2012sb}).} addressed so far is the impact of the choice of this input scale on the results. In the \emph{standard} approach followed by most groups the input scale is fixed at \emph{one} arbitrarily chosen value $Q_0^2 \geq 1$ GeV$^2$, and the consequences of the choice of a particular value (ranging from 1 to 9 GeV$^2$ in current analyses) are ignored.

Alternatively, the so-called \emph{dynamical} parton distributions \cite{Gluck:1994uf,Gluck:1998xa,Gluck:2007ck,JimenezDelgado:2008hf} are generated from input distributions at an \emph{optimally} determined low input scale $Q_0^2\equiv \mu^2<$ 1 GeV$^2$. This has the advantage that the input distributions \emph{naturally} tend to \emph{valence--like} (positive definite) functions, i.e., not only the valence but also the sea and gluon input densities vanish in the small--$x$ limit. Thus the behavior of the parton distributions at small $x$ appears in the dynamical approach as a consequence of QCD \emph{dynamics}, while in the \emph{standard} approach it has to be fitted. It was shown in \cite{Gluck:2007ck,JimenezDelgado:2008hf}, where also ``standard'' distributions were generated using $Q_o^2 = 2$ GeV$^2$, that this also implies that the dynamical distributions at {$Q^2$~\raisebox{-0.1cm}{$\stackrel{>}{\sim}$} 1 GeV$^2$ are more restricted and have smaller uncertainties than their standard counterparts. Note, however, that although the dynamical approach provides a natural connexion with non-perturbative physics, no claim is made that the dynamical distributions describe the nucleon at non-perturbative scales of typically $Q^2 \leq 1$ GeV$^2$; as a matter of fact no comparison with data is attempted for scales lower than the kinematic cuts defined below.

The choice of the input scale in a parton distribution analysis also affects the determination of QCD parameters. For example, the strong coupling constant \as is determined in \cite{Gluck:2007ck,JimenezDelgado:2008hf} together with the parton distributions and is substantially correlated with the gluon distribution which drives the QCD evolution (consequently its uncertainty is also smaller in the dynamical case.). At NNLO \cite{JimenezDelgado:2008hf} we got $\alpha_s(M_Z^2) =$ 0.1124 $\pm$ 0.0020 in the dynamical case, to be compared with $\alpha_s(M_Z^2) =$ 0.1158 $\pm$ 0.0035 in the ``standard'' one. These differences should be interpreted as genuine uncertainties in the determinations; e.g., our contribution to the next PDG value \cite{bethke} will be an average over dynamical and ``standard'' NNLO results, rather than a single particular value.

The crucial observation is that once one finds an optimal solution for the PDFs (and/or QCD parameters) using one input scale \qs, it is \emph{certain} that solutions exist at all other scales with equally good \chisq. As a matter of fact, one could find them via (forward or backward) RGE evolutions (this implies also that those solutions would have the same values of the QCD parameters, in particular of \as). As a consequence of this straightforward observation one can conclude that any dependence of the results on \qs is due to a certain \emph{inability} of the estimation procedure to find the optimal solution. For example one can immediately think of shortcomings of the parametrization to reproduce the optimal shape of the distributions at the different input scales, i.e. parametrization bias. The observation is however much more general and applies to the totality of the theoretical framework, as well as to the statistical estimation; we will refer to it as \emph{procedural bias}.

Turning the argument around, one can use the \qs dependence of the results as an \emph{estimator} of the bias due to the particular method used. Of course, since the \emph{true} solution is not known, it is not possible in general to determine the \emph{absolute} procedural bias. However variations with \qs give information on the its \emph{relative} size (i.e., with respect to the best solution found), and certainly set a lower limit on it, i.e., define a minimal error which, at the very least, has to be taken into account. Nevertheless, under the assumption that further improvements in the solution would be negligible, one could construct a measure of the procedural bias by quantifying variations of the results with \qs.

An additional observation is that for a significant \qs dependence to develop it is necessary to compare \qs values at which the shape of the distributions is quite different, e.g. a particular parametrization will adapt very similarly to distributions at input scales which are very close to each other. Note however that in the low \qs region the parton distributions functions go through a complete rearrangement, in particular the gluon and sea distributions change dramatically from a valence-like structure (or even negative values) to a more conventional \emph{standard} shape where they increase as $x$ decreases. Thus it is possible to estimate the procedural bias by studying the results obtained at different input scales in these two qualitatively differentiated regions, as was done in \cite{Gluck:2007ck,JimenezDelgado:2008hf}.

\begin{table}
\scriptsize
\begin{center}
\begin{tabular}{|c|c|c|c|c|}
\hline
                  &  NNLO13  &   NNLO17 &   NNLO20 &  NNLO22 \\
\hline
 $a_{u_v}$        & $\surd$  & $\surd$  & $\surd$  & $\surd$ \\
 $b_{u_v}$        & $\surd$  & $\surd$  & $\surd$  & $\surd$ \\
 $A_{u_v}$        & --       & --       & $\surd$  & $\surd$ \\
 $B_{u_v}$        & --       & $\surd$  & $\surd$  & $\surd$ \\
 $C_{u_v}$        & $\surd$  & $\surd$  & $\surd$  & $\surd$ \\
\hline
 $a_{d_v}$        & $\surd$  & $\surd$  & $\surd$  & $\surd$ \\
 $b_{d_v}$        & $\surd$  & $\surd$  & $\surd$  & $\surd$ \\
 $A_{d_v}$        & --       & $\surd$  & $\surd$  & $\surd$ \\
 $B_{d_v}$        & --       & --       & $\surd$  & $\surd$ \\
 $C_{d_v}$        & --       & --       & --       & --      \\
\hline
 $N_{\Delta}$     & $\surd$  & $\surd$  & $\surd$  & $\surd$ \\
 $a_{\Delta}$     & $\surd$  & $\surd$  & $\surd$  & $\surd$ \\
 $b_{\Delta}$     & $\surd$  & $\surd$  & $\surd$  & $\surd$ \\
 $A_{\Delta}$     & --       & $\surd$  & $\surd$  & $\surd$ \\
 $B_{\Delta}$     & --       & --       & --       & --      \\
 $C_{\Delta}$     & --       & --       & --       & --      \\
\hline
 $N_{\Sigma}$     & $\surd$  & $\surd$  & $\surd$  & $\surd$ \\
 $a_{\Sigma}$     & $\surd$  & $\surd$  & $\surd$  & $\surd$ \\
 $b_{\Sigma}$     & $\surd$  & $\surd$  & $\surd$  & $\surd$ \\
 $A_{\Sigma}$     & --       & $\surd$  & $\surd$  & $\surd$ \\
 $B_{\Sigma}$     & --       & --       & $\surd$  & $\surd$ \\
 $C_{\Sigma}$     & --       & --       & --       & --      \\
\hline
 $a_{g}$          & $\surd$  & $\surd$  & $\surd$  & $\surd$ \\
 $b_{g}$          & $\surd$  & $\surd$  & $\surd$  & $\surd$ \\
 $N^\prime_{g}$   & --       & --       & --       & $\surd$ \\
 $a^\prime_{g}$   & --       & --       & --       & $\surd$ \\
\hline
\end{tabular}
\caption{Definition of the different parametrizations used. A ``$\surd$'' symbol indicates that the parameter is set free in the corresponding parametrization, a ``--'' symbol that it is set to zero. Note that the parameters $N_{u_v}$, $N_{d_v}$ and $N_g$ are determined via quark--number and momentum conservation, respectively, and thus are not free.}
\label{parametrizations}
\end{center}
\end{table}

With the aim of investigating these ideas, we extend in the present paper the studies in \cite{Gluck:2007ck,JimenezDelgado:2008hf} to a systematic investigation on the effect of the input scale in the determination of parton distributions and QCD parameters (we will consider explicitly \as but most of the discussion applies equally to heavy quark masses). This studies are part of an ongoing update of the JR distributions, details on the NNLO framework can be found in \cite{JimenezDelgado:2008hf}, and changes for the update have been recently reported in \cite{DIS2012}. While most of the changes are irrelevant for the discussion here, some of them are related to the \as values obtained and will be discussed below. The exact procedure for statistical estimates is also of relevance and will be discussed in the Appendix.

In principle there is complete freedom for the choice of the scale at which the initial conditions for the RGE are taken. However, it is expected that non-perturbative dynamics become relevant and eventually dominant at low scales, e.g. scales of about 0.4 GeV$^2$ were estimated in \cite{Diakonov:1996sr} on the basis of chiral symmetry breaking. These are similar to (or lower than) the optimal scales considered in the dynamical determinations of parton distributions, which are determined on purely phenomenological grounds, i.e. the stability of the fits and a valence-like structure of the distributions \cite{Gluck:1994uf,Gluck:1998xa,Gluck:2007ck,JimenezDelgado:2008hf}. We will consider here fixed scales ranging from 0.6 GeV$^2$ to the 9 GeV$^2$ used by the ABM group \cite{Alekhin:2012ig}, which is the largest one used in current fits; in particular we have chosen $Q_o^2 = 0.6,\ 0.7,\ 0.8,\ 1,\ 2,\ 5,\ 9$ GeV$^2$.

For the purpose of comparing results with different degrees of procedural bias we consider parametrizations with increasing flexibility, ranging from 13 to 22 free parameters in addition to \as, which is also taken as a free parameter in all fits. The nomenclature and precise definition of each parametrization are given in Table \ref{parametrizations}. For the quark distributions  $u_v=u-\bar{u}$, $d_v=d-\bar{d}$, $\Sigma=\bar{u}+\bar{d}$, and $\Delta=\bar{d}-\bar{u}$ we use the following parametric form:
\begin{equation}
xf(x,Q_0^2) = N_f\ x^{a_f}(1-x)^{b_f}(1 + A_f \sqrt{x} + B_f x + C_f x^2 )
\end{equation}
were the polynomial coefficients are progressively set free. There are no free parameters for the strange-quark distributions \cite{DIS2012}. The deep-inelastic-scattering (DIS) and Drell-Yan data included in our analysis \cite{JimenezDelgado:2008hf,DIS2012} do not directly constrain the gluon distribution at high $x$, thus in order to increase the freedom of this distribution beyond the basic shape used in \cite{JimenezDelgado:2008hf} we consider
\begin{equation}
xg(x,Q_0^2) = N_g\ x^{a_g}(1-x)^{b_g} \left(1 + N^\prime_g\ x^{a^\prime_g}(1-x)^{25}\right)
\label{gluon}
\end{equation}
This shape allows the input gluon to become negative in the small $x$ region, a feature that the MSTW \cite{Martin:2009iq} group finds preferred by the HERA data if an input scale of $Q_o^2 = 1$ GeV$^2$ is used. It should be noted that due to the NNLO gluon--gluon splitting function, which is {\em negative} and {\em more} singular in the small--$x$ region \cite{ref26} than the LO and NLO ones, any NNLO gluon at low scales tend to decrease as $x$ decrease and might eventually turn negative. In particular at the low scales ($\simeq 0.6$ to 1 GeV$^2$) considered here, also our dynamical JR09 gluon \cite{JimenezDelgado:2008hf} exhibit this feature, even though it is generated from a definite positive input distribution at $Q_o^2 = 0.55$ GeV$^2$.

\begin{figure}
\centering
\includegraphics[width=0.75\textwidth]{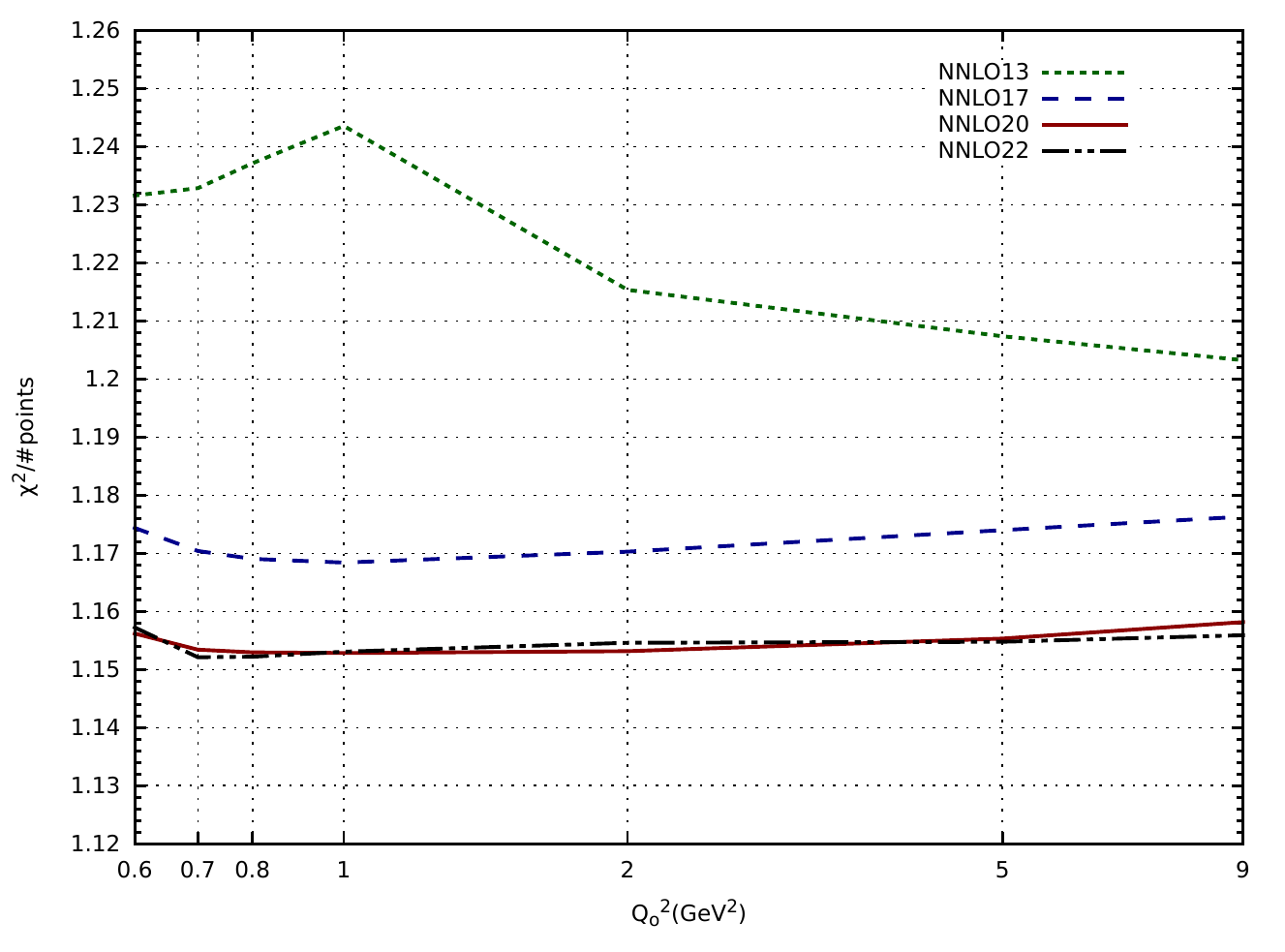}
\caption{\chisq values over number of points for 2411 points obtained for the different parametrizations using several input scales$^2$.}
\label{chi2_qs}
\end{figure}

In Fig.~\ref{chi2_qs} we present the \chisq (over number of points for 2411 points) values obtained at the different input scales for the parametrization considered in the analyses. The \chisq decrease as the number of free parameters increase until they stabilize at NNLO20, which indicate that the extra freedom of the second term in Eq.~\ref{gluon} does not improve \footnote{It might seem strange that the \chisq values for NNLO22 are slightly larger than for NNLO20 at some \qs values. This has to do with the rescaling of the systematic errors, as explained in the Appendix.} the description of the data achieved with the simpler parametrization with $N^\prime_g=0$ used in \cite{JimenezDelgado:2008hf}. It should be emphasized that for the NNLO22 fit with $Q_o^2 = 0.6$ GeV$^2$ the powers in  Eq.~\ref{gluon} turn out to be $a_g \simeq 1.2$ and $a^\prime_g \simeq 1$, i. e. even with this parametrization the input gluon distribution turn out to be valence--like as in \cite{JimenezDelgado:2008hf}. Thus, this is a
\emph{natural} tendency of the input gluons at low scales and \emph{not} an artifact of the simpler parametrization with $N^\prime_g=0$. As a matter of fact if one chooses a sufficiently low input scale the details of the input distribution at low $x$ do not have much influence in the predictions at higher scales, say $Q^2 > 2$ GeV$^2$, which is one of the aspects exploited in the dynamical approach to parton distributions.

Of more relevance for the present investigations is the fact that, as expected, the variation in \chisq among fits with different input scales also reduces as the flexibility of the parametrization increases; again the NNLO22 version does not reduce this (already rather small) variation further. One could (and should), of course, try to improve the theoretical description, by extending the parametrization or otherwise \footnote{Note as well that different methods to explore possible parametrization bias are currently in use, e.g. \cite{Ball:2011eq,Watt:2012tq}.}. However the suggestion in this article is that one can estimate the uncertainty due to the \emph{remaining} procedural bias by using this variation. This would lead to an additional theoretical error due to procedural bias which should be added (as an independent source, e.g. in quadrature) to the existing experimental and theoretical errors for both QCD parameters and predictions based on the extracted PDFs. As a matter of fact, one \emph{could} devise a quantitative measure of this error, say (half) the difference between the typical dynamical scale (about 0.6 GeV$^2$) and the most different ``standard'' one (say 2 GeV$^2$), instead of assuming it to be exactly zero.

\begin{figure}
\centering
\includegraphics[width=0.75\textwidth]{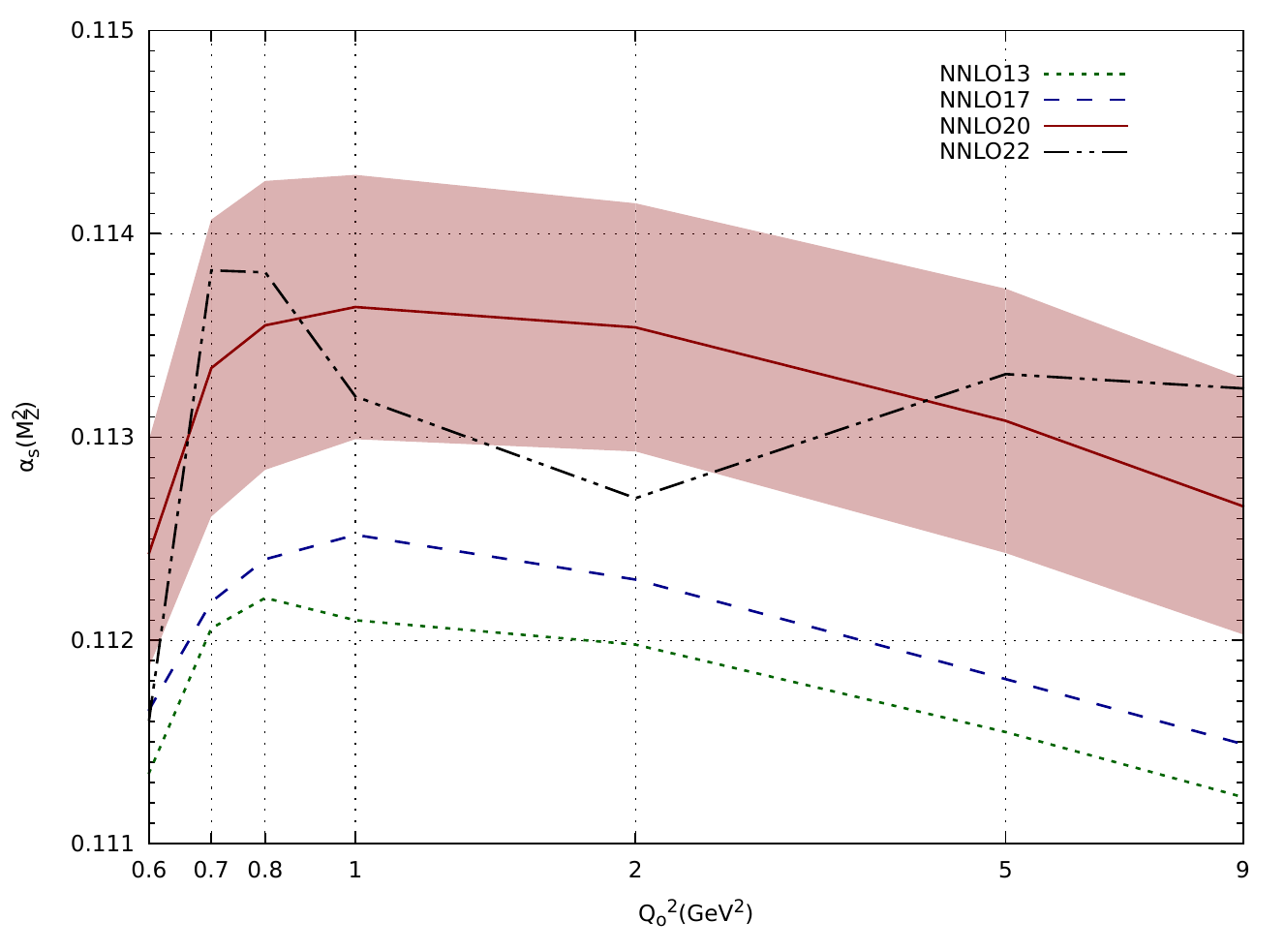}
\caption{\as values obtained for the different parametrizations using several input scales. The band indicates the error due to propagation of the experimental uncertainties for the NNLO20 results$^3$.}
\label{as_qs}
\end{figure}

Turning now to the determination of \as we present in Fig.~\ref{as_qs} our results for the different fits. It can be seen that the parametrizations with less number of free parameters lead to estimations which change as one increases the flexibility, while the results stabilize at NNLO20; further increasing the number of parameters does not lead to substantially different results but generates fluctuations. Note, however, that in the case of \as the variation with the input scale is not significantly reduced as the flexibility of the parametrization is augmented. Following the suggestion of the previous paragraph, an error of about 0.0006 due to procedural bias would be attributed to the determinations of \as in this analysis. Note that this is comparable to the uncertainty stemming from the experimental uncertainties\footnote{The uncertainty due to propagation of experimental errors reported here has been obtained using $\Delta \chi^2 = 1$.} of the data, which (depending on \qs) is also of about 0.0006, as indicated (for NNLO20 only) by the band in Fig.~\ref{as_qs}.

Although a full description of the ongoing update of our analyses will be published somewhere else \cite{preparation}, and quantitative results are not the main aim of this letter, the quoted values on \as are in need of some comments, in particular in relation to the results reported in \cite{JimenezDelgado:2008hf}. In addition to a full treatment of the experimental correlations, which is discussed in the Appendix, changes in our analysis framework include the fact that for the SLAC \cite{Whitlow:1991uw} and NMC \cite{Arneodo:1996qe} data we use now the directly measured cross-sections (instead of the extracted structure functions) since potential problems with the extractions in \cite{Arneodo:1996qe} have been pointed out \cite{Alekhin:2011ey}. A wealth of deuteron data have also been included \cite{Whitlow:1991uw,Arneodo:1996qe,Adams:1996gu,Benvenuti:1989fm}, for which description we use the nuclear corrections of \cite{Accardi:2011fa}. Further theoretical improvements include the use of the running mass
definition for DIS charm and bottom production and an approximated NNLO expression \cite{Alekhin:2010sv}, so that the semi-inclusive HERA data on heavy quark production included in our analysis up to NLO \cite{Gluck:2007ck} are now used at NNLO as well.

Another issue raised by the ABM collaboration is the necessity of including higher twist contributions in the description of fixed target data \cite{Alekhin:2012ig}, even if moderate kinematic cuts are used to select the data included in the fits, in particular for SLAC and NMC data \footnote{See also \cite{Thorne:2011kq,NNPDF:2011aa}.}. The kinematic cuts in our G(JR) analyses \cite{Gluck:2007ck,JimenezDelgado:2008hf} were $Q^2 \geq 4$ GeV$^2$ in virtuality and $W^2\geq 10$ GeV$^2$ in DIS invariant mass squared, and were applied to the $F_2$ values extracted from different beam energies and combined. The description was good for NMC data and rather poor for SLAC data. However since the number of points for these experiments was rather small (about 100 data points for NMC and 50 for SLAC), the values of the cuts did not affect much the results in \cite{Gluck:2007ck,JimenezDelgado:2008hf}.

This picture changes if, as the ABM collaboration does, data on the cross sections for individual energies are used, which amounts to hundreds of data points for each experiment. Since these data are also used in the present analysis, the virtuality cut has been raised to $Q^2 \geq 9$ GeV$^2$ for the SLAC and NMC data; the other data sets are less sensitive to higher twist \cite{Alekhin:2012ig}. We find results which are rather similar to our JR09 analysis \cite{JimenezDelgado:2008hf}, with $\alpha_s(M_Z^2)$ values about 0.113 to 0.114, depending on the input scale. If the usual cut of $Q^2 \geq 4$ GeV$^2$ is applied to these data $\alpha_s(M_Z^2)$ rises to 0.1176; a tendency in agreement with the ABM observations \cite{Alekhin:2012ig}. The inclusion of higher twist terms in the theoretical description should strongly reduce the dependence of the outcome of the fits and is currently under investigation \cite{preparation}. Yet another way of (wrongly) obtaining higher \as values is pointed out in the Appendix.

To summarize, we have presented a (first) systematic study of the effects of the choice of the input scale in global determinations of parton distributions and QCD parameters. It has been shown that although in principle the result should not depend on these choices, in practice a relevant dependence develops as a consequence of what we call \emph{procedural bias}. This uncertainty should be considered in addition to other theoretical and experimental errors, and a practical procedure for its estimation based on variations of the input scale has been proposed.

\section*{Acknowledgements}
We thank E.~Reya for a fruitful collaboration and for carefully reading the manuscript, and S.~Alekhin, J.~Bl\"umlein, S.~Moch, and V.~Radescu for discussions. This research is supported in part by the Swiss National Science Foundation (SNF) under contract 200020-138206. Authored by Jefferson Science Associates, LLC under U.S. DOE Contract No. DE-AC05-06OR23177. The U.S. Government retains a non-exclusive, paid-up, irrevocable, world-wide license to publish or reproduce this manuscript for U.S. Government purposes.

\section*{Appendix}
An improvement in our current analysis framework as compared to the published in \cite{JimenezDelgado:2008hf} is a complete treatment of the systematic uncertainties of the data including experimental correlations. The least-squares estimator that we use to take them into account has been explicitly written down in Appendix B of \cite{Stump:2001gu}; we repeat it here for convenience of the reader, and to fix the notation for further discussion. The global \chisq--function is obtained as the sum of the functions of each of the data sets included in the analysis. These consist of $i=1,\ldots,N$ data points of central value $D_i$, total uncorrelated error $\Delta_i$ (statistical and uncorrelated systematic errors added in quadrature), and correlated systematic errors $\Delta_{ij}$ for $j=1,\ldots,M$ sources. Denoting the respective theoretical predictions as $T_i$, the \chisq function for a data set is:
\begin{equation}
 \chi^2 = \sum_{i=1}^{N} \frac{1}{\Delta_i^2}\left( D_i + \sum_{j=1}^{M} r_j \Delta_{ji} - T_i \right)^2 + \sum_{j=1}^M r_j^2
\label{chi2def}
\end{equation}
In addition to the numerical minimization which determines optimal values for the parameters on which the theoretical predictions depend, one could also determine the  \emph{systematic shifts} $r_j$ via numerical optimization. However, since the dependence on this parameters is simple, it is also possible to do it analytically \cite{Stump:2001gu}
\begin{equation}
r_j = - \sum_{k=1}^{M} A_{jk}^{-1} B_k\, ,\hspace{1em} B_j = \sum_{i=1}^N\Delta_{ji}\frac{D_i-T_i}{\Delta_i^2} \, ,\hspace{1em} A_{jk} = \delta^K_{jk} + \sum_{i=1}^N \frac{\Delta_{ji}\Delta_{ki}}{\Delta_i^2}
\label{shifts}
\end{equation}
where $\delta^K_{jk}$ denotes here the Kronecker delta. Note that in this expression the errors have been regarded as independent quantities. This is usually true for the statistical (counting) errors, while the systematic errors are often proportional to the central values (typical examples are acceptance corrections and luminosity uncertainties), which requires a careful treatment.

Denoting \cite{Aaron:2009aa} the relative errors as $\delta = \frac{\Delta}{D}$ (for each point and all different kind of uncertainties), one would naively use $\Delta_{ji} = \delta_{ji}D_i$, \,$\Delta_i^2 = \delta_{stat,i}^2 D_i^2+ \delta_{unc,i}D_i^2$ in the above equations. However this is well-known to result in a biased estimation of the theory. First of all (even in the absence of correlations) data points with smaller central values wold have smaller uncertainties, leading to a bias towards small theoretical predictions. Moreover, when correlations are taken into account, points with larger central values would be shifted (for a particular $r_j$) more than those with smaller ones, thus resulting in smaller theoretical predictions (following the smallest central values) and larger systematic shifts (to accommodate the points with larger ones). This effect is very important for normalization uncertainties, leading to shifts of several (negative) units and clearly biased estimators  (see also \cite{Ball:2009qv} for some simple illustrations).

A solution to this problem is to take the absolute systematic errors proportional to theoretical predictions instead of proportional to the data. Note, however, that merely setting $\Delta_{ji} = \delta_{ji}T_i$, \, $\Delta_i^2 = \delta_{stat,i}^2 D_i^2+ \delta_{unc,i}T_i^2$ in Eqs.~\ref{chi2def}--\ref{shifts} \label{chi2def,shifts} also leads to biased results, in this case towards larger theory values since it would be possible to decrease \chisq by merely increasing the theory estimations, thus increasing the systematic errors and decreasing the shifts. An unbiased solution can be found iteratively: one \emph{scales} the errors as above using a (fixed) initial theory $T = T^{(0)}$, finds a first ``optimal'' solution $T^{(1)}$ and proceed to update the errors by rescaling them using $T=T^{(1)}$ as fixed theory, thus iterating the process until the result is stable, say $T^{(n)} = T^{(n-1)}$ up to the desired precision. In practice the algorithm converges very rapidly (already at the second/third iteration). Note that this is the method used for the combination of inclusive HERA data \cite{Aaron:2009aa}.

\begin{figure}[t]
\centering
\includegraphics[width=0.6\textwidth]{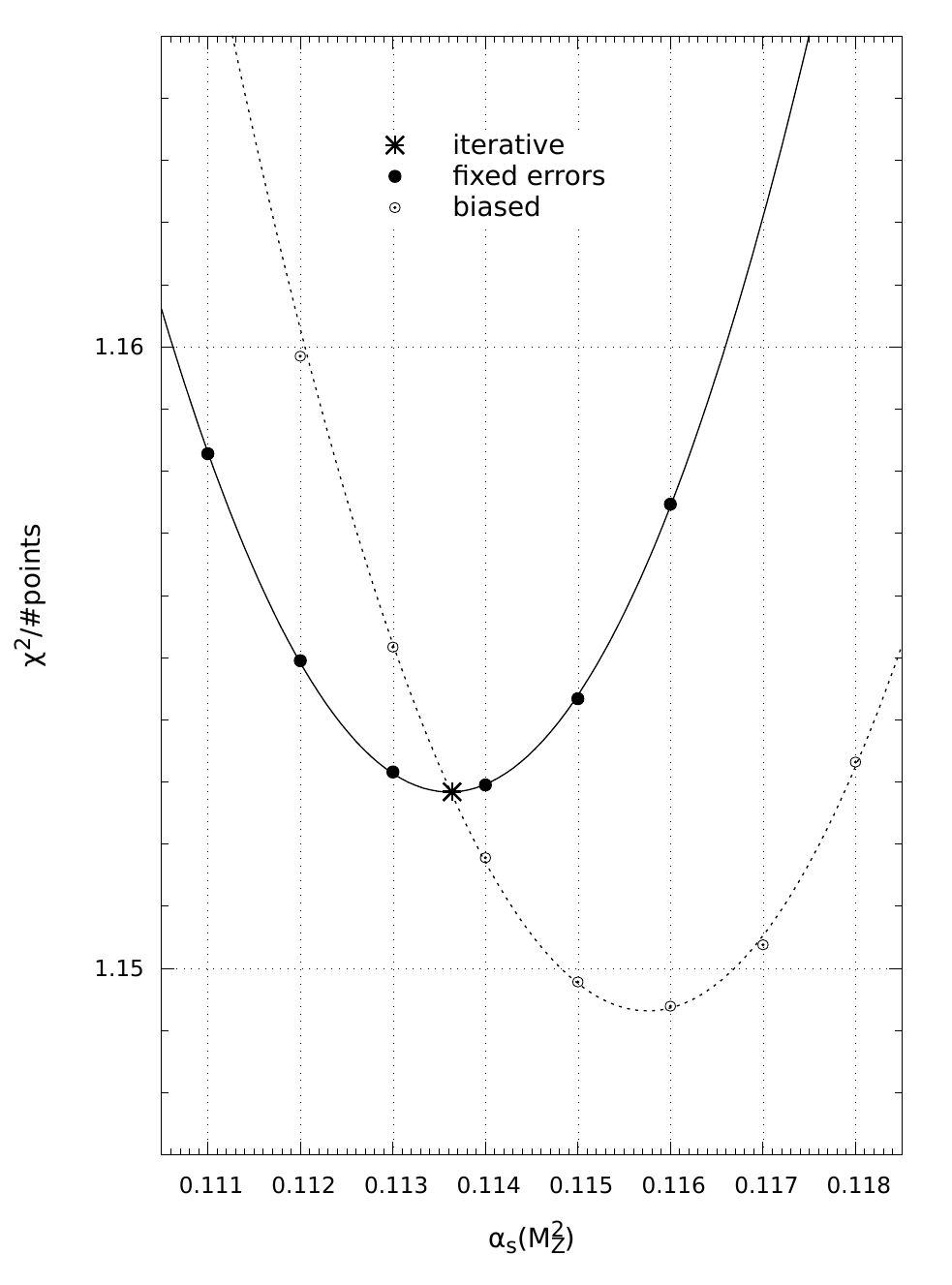}
\caption{\as scans for our NNLO20 parametrization at $Q_o^2 =$ 1 GeV$^2$. The iterative result is compared with the estimation from a scan with fixed errors (obtained by using the iterative result for scaling), and with the biased result that would be obtained by continuously rescaling the systematic errors (biased).}
\label{scan}
\end{figure}

A note of warning. If one uses this procedure for \chisq in parameter scans, i.e. to determine the \chisq profile by fits at fixed values of a parameter, one encounters again a bias towards larger theory estimators, for the same reasons as before (\chisq is smaller for the parameter values which produce larger systematic errors). However, this kind of scans can lead to a sensible (unbiased) value for the scanning parameter if a \emph{fixed} theory is used for the rescaling; in this case the \chisq values depend on the chosen theory but are at least comparable. The situation has been illustrated in Fig.~\ref{scan}, where we present a \as scan obtained with a fixed theory (NNLO20 at $Q_o^2 = 1$ GeV$^2$ from the main text), and compare it with what is obtained by continuously rescaling the systematic errors. This could be one of the reasons for the higher \as values reported by some PDF groups, e.g. \cite{Radescu:2011cn} and/or \cite{Ball:2011us}.

\end{document}